\documentclass{aa}  
\usepackage{graphicx}
\usepackage[varg]{txfonts}
\usepackage{natbib}
\bibpunct{(}{)}{;}{a}{}{,} 
\begin{document}
   \title{A new \textit{Herschel} view of the young star T54:\\not a transitional disk?\thanks{\textit{Herschel} is an ESA space observatory with science instruments provided by European-led Principal Investigator consortia and with important participation from NASA.}}


     \author{L.~Matr\`{a}\inst{\ref{inst1}, \ref{inst2}}
   		\and
   			B.~Mer\'{i}n\inst{\ref{inst1}} 
	\and
          C. Alves de Oliveira\inst{\ref{inst1}}
          \and
          N. Hu\'elamo\inst{\ref{inst3}}
          \and
	A.~K\'osp\'al\inst{\ref{inst4}} 
           \and
	N.~L.J.~Cox\inst{\ref{inst5}}
          \and
          \'A.~Ribas\inst{\ref{inst1}, \ref{inst3}}
          \and
          E.~Puga\inst{\ref{inst1}}
	\and
          R.~Vavrek\inst{\ref{inst1}} 
          \and
	P.~Royer\inst{\ref{inst5}} 
         	\and 
          T.~Prusti\inst{\ref{inst4}} 
          \and
	G.~L.~Pilbratt\inst{\ref{inst4}} 
	\and
	P.~Andr\'e\inst{\ref{inst6}} 
%
          }


	\institute{
	Herschel Science Centre, ESAC-ESA, P.O. Box, 78, 28691 Villanueva de la Ca\~{n}ada, Madrid, Spain\label{inst1} \\
              \email{lmatra@sciops.esa.int}
          \and
      School of Physics, Trinity College Dublin, Dublin 2, Ireland\label{inst2} 
         \and
        Centro de Astrobiolog\'{i}a, INTA-CSIC, P.O. Box - Apdo. de correos 78, Villanueva de la Ca\~nada Madrid 28691, Spain\label{inst3} 
          \and
        Research and Scientific Support Department, ESTEC-ESA, PO Box 299, 2200 AG Noordwijk, The Netherlands\label{inst4} 
         \and
        Instituut voor Sterrenkunde, K.U. Leuven, Celestijnenlaan 200D, B-3001, Leuven, Belgium\label{inst5} 
        \and
        Laboratoire AIM Paris -- Saclay, CEA/DSM -- CNRS -- Universit\'e Paris Diderot, IRFU, Service d'Astrophysique, Centre d'Etudes de Saclay, Orme des Merisiers, 91191 Gif-sur-Yvette, France\label{inst6} 
              }

   \date{Received -; Accepted 25 October 2012}

 
  \abstract
   {Observations of transitional disks give us an understanding of the formation of planets and planetary systems such as our own. But care must be taken in the identification of such sources: the higher spatial resolution of the  \emph{Herschel} Space Observatory provides a new view on the origin of the far-infrared and sub-millimeter excesses observed.}
   {We review the nature of previously known transitional disks in the Chamaeleon I star-forming region with \emph{Herschel} data.} 
   {We analyze \emph{Herschel} PACS and SPIRE images of the young star T54 together with ancillary images. We also analyze its spectral energy distribution and indications from optical and mid-infrared spectroscopy.}
   {We detect extended emission in the PACS 70 $\mu$m image $\sim$6$''$ off source at a position angle of 196$^{\circ}$ from T54. The emission detected at longer wavelength (PACS 100, 160, SPIRE 250 and 350 $\mu$m) is also offset from the position of the star. This suggests that the excess observed in the far-infrared part of the SED is not fully associated with T54. }
   {\emph{Herschel} images show that the far-infrared excess seen in T54 is not due to a transitional disk but to extended emission south-west of the source. The object still shows point-like and now downscaled excess at mid-infrared wavelengths, but its origin cannot be constrained without higher spatial resolution data. However, different indications point towards an evolved disk or extended unresolved emission close to the source.}


\maketitle

%

\section{Introduction}

Transitional disks (TDs) are circumstellar disks that exhibit inner clearings or gaps induced by physical processes such as photoevaporation, grain growth, and dynamical interactions with stellar companions or candidate planets. This causes a wide range of spectral energy distribution (SED) properties to be observed \citep[][and references therein]{Williams2011}. Disks are identified as transitional if they have no or small near-infrared excess, steep slopes in the mid-infrared, and large far-infrared excesses \citep[e.g.][]{Merin2010}. However, different definitions used sometimes make their identification problematic. In fact, similar SEDs can be reproduced by a number of environments, including background objects and nebulosity in the source surroundings, which can contaminate and bias the sample of known transitional disks. In addition to this, asymptotic giant branch stars and classical Be stars can easily be mistaken as TDs with significant flux deficit at all wavelengths and $\alpha_{excess}$$<$0 \citep[e.g.][]{Cieza2010}, and SEDs of edge-on protoplanetary disks can look like those of TDs with a sharp rise in the mid-IR \citep{Merin2010}. See review from \cite{Williams2011} and references therein for a more detailed description. \\
\indent In order to accomplish a more thorough characterization of this class of young objects, there is a need to examine their fluxes at far-infrared wavelengths. \emph{Herschel}, with its improved spatial resolution, can greatly serve the purpose of ruling out sources affected by contamination. In the Chamaeleon I star-forming region (Cha I), the young object T54 (also \object{NAME HM Anon}) is one of 8 candidate transitional disks \citep[][]{Manoj2011}. The star has been reported as a spectral type G8 \citep{Luhman2007}, weak-lined T Tauri (WTTS) \citep{Nguyen2012}, with visual extinction of 1.78 mag and luminosity of 4.1 L$_{\odot}$ \citep{Kim2009}. 
T54 is a known subarcsecond binary \citep{Ghez1997,Lafreniere2008}. The companion is located at a projected separation of 0$\farcs$247 (43\,AU) and position angle (PA) of $246.5^{\circ}$. Resolved optical spectra by \citet{Nguyen2012} suggest that the optical luminosity of the system is dominated by the primary component, although we cannot rule out that the IR excess at the position of T54 discussed later is partially originated from the secondary.


\section{Observations}\label{Observations}

\subsection{Herschel}

The Chamaeleon I region was observed by the \emph{Herschel} Space Observatory \citep{Pilbratt2010} as part of the Gould Belt Survey \citep{Andre2010}. Detailed description of the observations can be found in \citet{Winston2012}. Observations used in this paper have obsids 1342213178, 1342213179 (22 Jan. 2011) for parallel-mode PACS \citep{Poglitsch2010} 70 and 160 $\mu$m and SPIRE \citep{Griffin2010} 250 and 350 $\mu$m bands. Additional PACS observations were obtained at 100 $\mu$m (obsids 1342224782, 1342224783, 1342225003, and 1342225004; dates 27 Jul. 2011 and 1 Aug. 2011). The data were reduced using HIPE \citep{Ott2010} version 8.2.0 and processed using the Scanamorphos software v14 \citep{Roussel2012} for PACS, and the Scan Map Destriper pipeline in HIPE for SPIRE. The astrometry of the PACS images was refined using 2MASS PSC positions of nearby point sources through the Astrometrical Calibration tool in the Aladin v7 software \citep{Aladdin}.

Photometry was extracted using the HIPE \emph{annularSkyAperturePhotometry} task, which performs aperture photometry with background subtraction. For PACS, aperture corrections were applied and photometric errors were estimated as specified in the PACS Point-Source Flux Calibration Technical Note from April 2011. For SPIRE, the same was applied using apertures and corrections from Section 5.7.1.2 of the SPIRE Data Reduction Guide, with a conservative uncertainty of $\pm$10\%.

\subsection{Ancillary data}

Ancillary images were retrieved in order to check, through visual inspection, the possible presence of counterparts for the extended emission around T54 discovered by \emph{Herschel}. 
The FORS 1 H$\alpha$ image (PROG ID 075.C-0809(A),  26 Jul. 2005) was obtained from the \emph{ESO Archive}, while 2MASS \emph{All-Sky Release Survey} images (25 Jan. 2000) were obtained from the 2MASS \emph{Interactive Image Service}. IRAC (AOR key 20014592, 16 May 2007) and MIPS (AOR key 5706752, 24 Dec. 2009) Post BCD images were obtained from the \emph{Spitzer Heritage Archive}. The IRS spectrum (AOR key 12695552) was extracted from the \emph{Spitzer Heritage Archive} and reduced with SPICE v2.5.0. The short low slit (5.2--14.5$\,\mu$m) included only T54, while the long low slit (14.0--38.0$\,\mu$m) included both T54 and 2MASS J11124076-7722378. While the PSFs slightly overlap, the sources are resolved enough that an uncontaminated spectrum of T54 could be extracted by using a narrow aperture. 
 
\section{Results}

\begin{figure*}
  \centering
   \includegraphics[width=1.00\hsize]{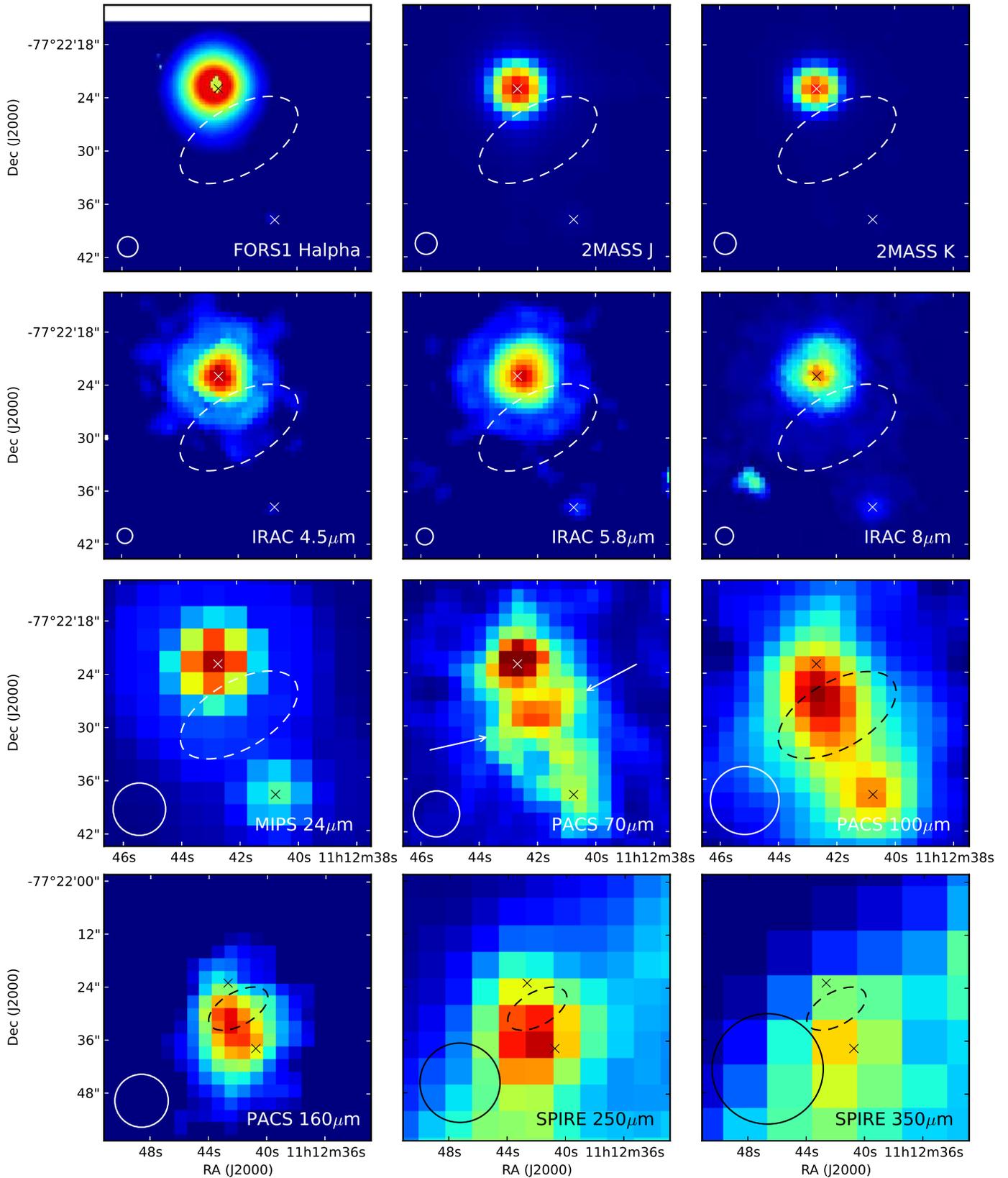}
   \caption{Images of T54 at different wavelengths. North is up, east is left. The bottom row images show a region twice as large as the other ones. The diameter of the circles in the bottom left of each image represents the FWHM of the PSF for the different observations. The northern cross is centered at the position of T54 from the 2MASS \emph{Point Source Catalog}, while the southern one corresponds to the position of 2MASS J11124076-7722378. Arrows indicate the location of extended emission as detected at 70 $\mu$m, and the dashed ellipse represents the same in all other images. Non-labelled objects represent artifacts. \label{fig:1}}
\end{figure*}

Figure \ref{fig:1} presents \emph{Herschel} and ancillary images centered at $RA_{2000} = 11^h12^m42 \fs029$ $Dec_{2000} = -77^{\circ}22'28\farcs58$ to display the source and its surroundings. 
T54 is clearly detected as a point source (top-left cross in Fig. \ref{fig:1}) at all optical, near- and mid-IR wavelengths. Extended emission is discovered in the PACS 70 $\mu$m image at a distance of $\sim$6$''$ and PA of 196$^{\circ}$ from T54. In the 100 $\mu$m image we observe emission in an elongated shape whose photocenter is offset from the star. This off source emission observed is unique to T54: it was not found in any \emph{Herschel} images of the other known transitional disks in the Chamaeleon I and II regions (Ribas et al., in prep.). In 160, 250 and 350 $\mu$m images, we also observe extended emission centered off source. However, we note that the increased PSF size makes it impossible to rule out a possible contribution from the nearby point source 2MASS J11124076-7722378 (bottom-right cross in Fig. \ref{fig:1}) at PA 203$^{\circ}$ and distance of 16$\arcsec$ from T54. Inspection of 2MASS color-color and color-magnitude diagrams shows very extreme colors compared to young Cha I members, indicating that this is likely an unrelated background object. No detection was obtained in SPIRE 500 $\mu$m, due to the presence of strong emission from the environment of the Cha I cloud.

\begin{table}
\caption{\emph{Herschel} photometry}
\label{tab:1} 
\centering
    \begin{tabular}{l c c c c c c c c c c c c c c c}
      \hline
      \hline\rule{0mm}{3mm} Instrument & Wavelength ($\mu$m) & Flux (Jy) & Aperture radius ($\arcsec$)  \\
      \\\hline\rule{0mm}{3mm} PACS & 70 (star-only) & 0.23$\pm$0.03 & 4 \\     
      \rule{0mm}{3mm} PACS & 70 (offset em.-only) & 0.22$\pm$0.03 & 4 \\
      \rule{0mm}{3mm} PACS & 70 (star+offset em.) & 0.48$\pm$0.03 & 8 \\
      \rule{0mm}{3mm} PACS & 100 & 0.61$\pm$0.02 & 8 \\
      \rule{0mm}{3mm} PACS & 160 & 1.10$\pm$0.11 & 22 \\
      \rule{0mm}{3mm} SPIRE & 250 & 0.49$\pm$0.05 & 22 \\
      \rule{0mm}{3mm} SPIRE & 350 & 0.24$\pm$0.02 & 30 \\
      \hline\rule{0mm}{3.0mm}
    \end{tabular}     
\end{table}

Table \ref{tab:1} presents \emph{Herschel} aperture corrected fluxes, together with the apertures used. Background sky annuli with radii of 25$\arcsec$ and 40$\arcsec$ for all PACS images, and 60$\arcsec$ and 90$\arcsec$ for SPIRE were subtracted.

\begin{figure}
   \resizebox{\hsize}{!}{\includegraphics{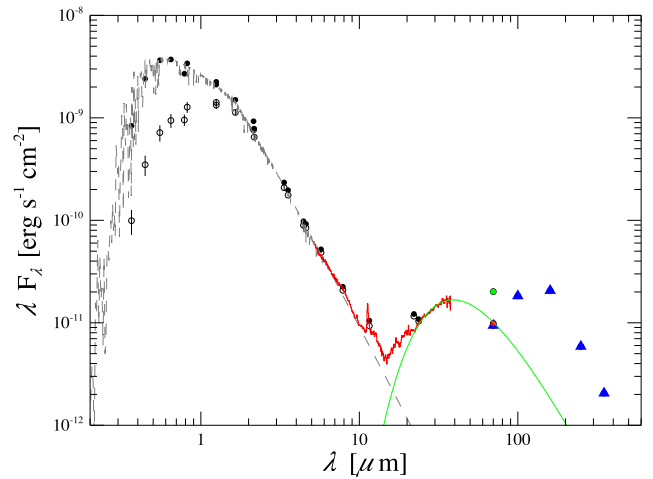}}
      \caption{Spectral energy distribution of T54. Open circles are observed optical, DENIS, 2MASS, WISE PRC, \textit{Spitzer} IRAC and MIPS 24 $\mu$m fluxes from the literature. Solid black circles are the fluxes dereddened with an A$_V$ of 1.78 mag using the law from \cite{Weingar2001}. The red line is the \textit{Spitzer} IRS spectrum. The dashed line is the G8 NEXTGEN stellar model that best fits the optical photometry. At 70 $\mu$m, PACS fluxes displayed are from the star only (red solid circle), from the extended emission only (blue solid triangle) and from both (green solid circle). Longer wavelength \emph{Herschel} flux values, not originated from the star position, are shown as blue solid triangles. The green solid line represents a T = 94 K blackbody curve fit to the mid-IR fluxes.  \label{fig:2}}
\end{figure}

Figure \ref{fig:2} presents the SED of the object updated with the values from Table \ref{tab:1}. We display dereddened fluxes obtained from the literature, and a best-fit NEXTGEN stellar model \citep{Hauschildt1999,Allard2000} for spectral type G8. The SED shows no or very little infrared excess at wavelengths $\lesssim$10 $\mu$m. The mid-IR excess at wavelengths up to 70 $\mu$m (red solid circle) is attributed to the unresolved circumstellar environment at the star position only, as confirmed by the \textit{Spitzer} MIPS 24 $\mu$m image.  The green solid circle in the SED represents the PACS 70 $\mu$m flux from both the extended emission observed in the corresponding image (Fig. \ref{fig:1}) and from the star position. At longer wavelengths we cannot attribute the measured fluxes to the star, but mostly to the nearby extended emission and possible additional contamination from 2MASS J11124076-7722378 (triangles in Fig. \ref{fig:2}). Hence, a significant part of the far-infrared excess seen in the SED of T54 (about half at 70 $\mu$m, likely more at increasing wavelengths) does not originate from the star position, but from nearby extended emission resolved by \emph{Herschel}.

If we assume that the remaining downscaled excess arises from thermal emission from dust grains, a single-temperature blackbody fit yields a dust temperature of approximately 94 K. We estimate the luminosity of the star by integration of the photospheric fluxes as $\sim$3.66 L$_{\odot}$, and the luminosity of the dust from integration under the mid-IR blackbody curve. This leads to a fractional luminosity value of L$_{IR}$/L$_{\ast}$ = 0.005.

We finally use the pre-main sequence tracks from \citet{Palla1999} in the census of Chamaeleon I by \citet{Luhman2004} to estimate the age and mass of T54. We find that the system is between 3 and 10 Myr old, with a mass of approximately 1.5 M$_{\odot}$. This is consistent with the age spread in the Chamaeleon I region found by \citet{Luhman2004}, in which objects of different mass range between 1 and 10 Myr of age. However, they also find a median age of $\sim$2 Myr, implying that T54 has a more evolved nature compared to most sources in the cluster.

\section{Discussion}
\label{discussion}

In this section we discuss previous informations about T54 from the literature in the context of the new \textit{Herschel} data. The main result is the discovery of extended emission contaminating the far-infrared part of the SED (discussed in Sect. \ref{4.1}), which scales down the total excess attributed to the source significantly. This potentially changes the nature of T54: in Sect. \ref{4.2} we draw a comparison with similar objects, and then conclude in Sect. \ref{4.3} by considering the new characteristics of the system if a disk is still to be present.

\subsection{Far-infrared extended emission}
\label{4.1}

In the PACS 70 $\mu$m image we clearly resolve extended emission with its flux peaking at a projected distance of 1040 AU and PA of 196$^{\circ}$ from the star. At longer wavelengths this emission is not resolved, but most of the flux comes from its position. Presence of nebulous emission has been reported in the literature. Both \citet{Gauvin1992} and \citet{Spangler2001} report the emission from the source being extended at IRAS and ISO wavelengths. Furthermore, the presence of a reflection nebula (referred to as \object{GN11.11.2} or \object{BRAN341E}) was reported in a study by \citet{Brand1986} and its position is consistent with the emission discovered. We conclude that about half of the excess at 70 $\mu$m is not to be associated with a transitional disk at the star position but to extended emission offset from it, likely associated with a small reflection nebula. At longer \emph{Herschel} wavelengths the latter is not resolved, but most of the excess seem to originate from it, and not from the position of the system. These results cause a reduction of the total excess flux attributed to T54 and thus modify the view we had of its nature.

\subsection{Comparison with other objects}
\label{4.2}

The system is classified as non-accreting due to its H$\alpha$ line in absorption \citep{Nguyen2012, Feigelson1989, Walter1992}, with the latter, however, indicating substantial filling in of the line, as in the case of the young object DoAr 21 \citep{Jensen2009}, discussed further below. This is in contrast with all but two of the rest of transitional disks in Chamaeleon I \citep{Manoj2011}. On the other hand, we note that \cite{Kim2009} report the source to be accreting from analysis of its \emph{U}-band flux. High H$\alpha$ and \emph{U}-band variability has been reported in objects of similar nature, such as T Cha \citep{Schisano2009} and DoAr21 \citep{Jensen2009}, even on a timescale of days. In addition to this, flaring activity in T54 has been suggested in the X-ray study by \citet{Feigelson1993} to explain the significant increase in flux between the \textit{ROSAT} and the previous \textit{Einstein} observations \citep{Feigelson1989}. This again enhances the similarity with DoAr 21, for which \citet{Jensen2009} suggested that flares could account for the U-band excess and variability observed. In summary, most references point towards a non-accreting stellar environment, but no conclusions can be drawn without further spectral observations.

T54 also presents polycyclic aromatic hydrocarbon (PAH) emission at 11.3 $\mu m$, which is undetected in the majority of disks around T Tauri stars \citep{Geers2006}. In this context, it is useful to compare T54 to the case of DoAr 21, since the latter is similarly a late spectral type with PAH features and lack of silicates. Interestingly, \citet{Jensen2009} obtained narrow-band images centered on the 11.3 $\mu m$ feature and found a partial arc or ring of dust at a projected distance of 134 AU from the source.
Hence, even in the case of T54, it is possible that the PAH emission originates from an extended area and is not associated with a circumstellar disk.

We also compare the X-ray properties of T54 to those of DoAr 21, and analyse how these can influence the presence of PAH emission. In the case of T54, \citet{Feigelson1993} report considerable X-ray luminosity (log(L$_X$) = 30.5 erg s$^{-1}$), which is not as extreme as the value for DoAr 21 \citep[log(L$_X$) $\simeq$ 32,][]{Jensen2009}. \citet{Jensen2009} suggest that strong X-ray emission could be responsible for exciting the PAHs. However, a more recent study by \citet{Siebenmorgen2010} shows that X-rays destroy PAH molecules efficiently at all distances. This would make the detection of PAHs around such strong X-ray emitters very unlikely, and therefore cannot explain the 11.3 $\mu$m feature observed.

Furthermore, T54 does not display a 10 $\mu$m silicate feature even though transitional disks commonly show it \citep{Manoj2011}. A broad study of disks with inner holes by \cite{Merin2010} presents only three objects lacking this feature, namely DoAr 21, SSTc2d J18285808+0017243 and Sz 84, which interestingly also have SEDs very similar to that of T54, and PAH features in two of them. These three out of a sample of 35 disks were all classified as probable extended sources and therefore dubious transitional disks. However, we cannot exclude the presence of a disk with a very clean inner hole, such as T25 in Cha I \citep{Kim2009} and IRAS 04125+2902 in Taurus \citep{Furlan2011}. Therefore, the lack of the 10 $\mu$m silicate feature also raises doubts about the disk nature of the source.

\subsection{Concluding remarks}
\label{4.3}

If the downscaled excess, however, originated from a disk, this would have an evolved nature. In fact, its fractional luminosity of 0.005 is consistent with the definition of debris disk from \citet[][]{Wyatt2008} (L$_{IR}$/L$_{\ast}$ $<$ $10^{-2}$) and our age estimate sets the system among the most evolutionary advanced in the Cha I region. Using equation 3 in \cite{Wyatt2008}, we obtain a disk radius of 16.8 AU, which is smaller than the 43 AU binary separation \citep{Lafreniere2008} and would make T54 an interesting case of an evolved disk around a single binary component. However, no conclusions can be drawn without observations at higher spatial resolution, needed to better constrain the nature of this system.

\begin{acknowledgements}
We thank the referee for the useful comments and the suggested improvements to the structure of the paper.
This work has been possible thanks to the support from the ESA Trainee and ESAC Space Science Faculty and of the Herschel Science Centre. Support was also provided by the Enterprise Ireland Space Education Program, and by the Belgian Federal Science Policy Office via the PRODEX Programme of ESA.
This work is based in part on observations made with the Spitzer Space Telescope, which is operated by the Jet Propulsion Laboratory, California Institute of Technology under a contract with NASA; on data obtained from the ESO Science Archive Facility under request number MATRA 35567; and on data products from the Two Micron All Sky Survey, which is a joint project of the University of Massachusetts and the Infrared Processing and Analysis Center/California Institute of Technology, funded by the National Aeronautics and Space Administration and the National Science Foundation.
This research has made use of the SIMBAD database, operated at CDS, Strasbourg, France and of NASA's Astrophysics Data System.
\end{acknowledgements}
\bibliographystyle{aa}
\bibliography{biblio}


\end{document}